\documentclass[11pt,a4paper]{article}
\usepackage{cite,indentfirst}
\usepackage{amssymb,amsmath}
\usepackage[dvips]{epsfig}
\usepackage{epsf}
\usepackage{graphics}

\def\mprp{\mbox{\tiny $\bot$}}
\def\mprl{\mbox{\tiny $\|$}}

\def\beq{\begin{eqnarray}}
\def\eeq{\end{eqnarray}}
\def\ee{\varepsilon}
\def\lm{\lambda}

\newcommand{\prl}[1]{#1_{\mbox{\tiny $\|$}}}

\def\1{1 \to 1 \, 2}
\def\2{1 \to 2 \, 2}

\def\P{{\cal P}}

\title{\vspace*{-20mm}
\begin{flushright}
{\normalsize Yaroslavl State University\\
             Preprint YARU-HE-06/04\\
             hep-ph/0609192} \\[10mm]
\end{flushright}
{\bf The Compton effect  in  strongly magnetized  plasma}}

\author{M. V. Chistyakov, D. A. Rumyantsev \\[3mm]
{\small\it Division of Theoretical Physics,} \\
{\small\it Yaroslavl State (P.G.~Demidov) University,} \\
{\small\it Sovietskaya 14, 150000 Yaroslavl, Russian Federation}\\
{\small\tt E-mail: mch@uniyar.ac.ru, rda@uniyar.ac.ru}
}

\date{}

\begin{document}

\maketitle

\thispagestyle{empty}

\begin{abstract}

\baselineskip=22pt The process of Compton scattering $\gamma e^{\pm}
\to \gamma e^{\pm}$ in  strongly magnetized hot electron-positron plasma is considered. 
The analytical expressions for the partial cross sections in rarefied plasma are obtained.
The numerical estimations of the absorption rates for various scattering channels 
are  presented  with taking into account of the photon dispersion and wave function renormalization 
in  strong magnetic field and plasma. The comparison of the scattering absorption rate with photon splitting probability
shows the existence of the temperature range where these values are comparable with each other. 
The astrophysical applications of the obtaining results are discussed.
\end{abstract}

\section{Introduction}

\unitlength 1mm
Recent observations~\cite{Kouveliotou:1998ze,Kouveliotou:1998fd,Gavriil:2002mc,Ibrahim:2002zw} give ground to believe that some astrophysical objects (SGR and AXP) are magnetars,
a distinct class of isolated neutron  stars with magnetic field strength of
$B \sim 10^{14}-10^{16}$ G~\cite{Duncan:1992,Duncan:1995,Duncan:1996}, i.e. $B \gg B_e$, where
$B_e = m^2/e \simeq 4.41\times 10^{13}$~G~\footnote{We use natural units
$c = \hbar = k = 1$, $m$ is the electron mass,  $e > 0$ is the elementary
charge.} is the critical magnetic field.
The spectra analysis of these objects is also providing evidence for the
presence of electron-positron plasma in magnetar environment. It is
well-known that strong magnetic field and/or plasma could influence
essentially on different quantum processes~\cite{Duncan:2000pj,Lai:2001,KM_Book,Harding:2006qn}. Therefore the effect of  magnetized plasma on microscopic physics should be incorporated in the magnetosphere models of strongly magnetized neutron stars.

The various studies indicate that electromagnetic processes such as 
Compton scattering and photon splitting $\gamma \to \gamma \gamma$ (merging $\gamma \gamma \to \gamma$) 
could play a crucial role in these models. For example, the process of photon splitting in strong magnetic field could 
suppress  the production of electron-positron pair required for a 
detectable radio emission from magnetar~\cite{BH:1998, BH:2000}. In turn, Compton 
scattering may play an important role in the models of atmosphere emission 
from strongly magnetized neutron stars~\cite{Miller:1995, Bulik:1997}. 
The both processes are very important in the models the radiation emission from 
SGR burst~\cite{Duncan:1995, HBG:1996, HBG:1997}.

Previously, Compton scattering was studied under various  conditions. The 
case of the magnetic field without plasma was investigated 
in~\cite{Herold:1979, Melrose:1983, Harding:1986, Harding:2000}. It was found that Compton scattering  
becomes strongly anisotropic and essentially depends on photon 
polarization. In~\cite{Bulik:1997} it was shown that the 
dispersion properties of photon in  strongly magnetized cold plasma could 
significantly influence on the process under consideration.

In the present work the process of photon scattering, $\gamma e^{\pm} \to \gamma
e^{\pm}$, is investigated in the presence of strong magnetic field and
charge-symmetric electron positron plasma,
when the magnetic field strength $B$ is the maximal physical
parameter, namely $\sqrt{eB} \gg T,\, \omega, \, E$. Here $T$ is
the plasma temperature,  $\omega$ and $E$ is the initial photon and electron
energies. In this case almost all electrons and positrons in plasma are on the
ground Landau level.
We also consider the none resonance case, when electron and photon momenta
satisfy the condition $eB \gg (pk)$.
Such conditions could be realized e.g. in the Thompson \& Duncan model of SGR 
burst~\cite{Duncan:1995} in which magnetically trapped high temperature ($\sim 1$ 
MeV) plasma fireball could be created.

The main purposes of our work are:
\begin{itemize}
\item
to investigate the influence of the strongly magnetized hot plasma on the 
process of Compton scattering taking account of photon dispersion and wave 
function renormalization;
\item
to compare the Compton scattering with the process of photon splitting;
\item
to estimate the possible effect of the process under consideration on 
spectral formation of SGR.
\end{itemize}

The plan of the paper is following. In the section 2 we consider the 
propagation of the electromagnetic radiation in magnetized medium and give 
the expression of Compton scattering amplitudes for different polarization 
configurations of the initial and final photons. The analytic and numerical calculations of the 
photon absorption rate and cross section of the process are 
presented in Section 3. The discussion of the possible astrophysical 
application of Compton scattering are given in Section 4.  

\section{Photon propagation in strongly magnetized hot plasma}

\label{Sec2}

The propagation of the electromagnetic radiation in any active medium is
convenient to describe in terms of normal modes (eigenmodes). In turn, the
polarization and dispersion properties of normal modes are connected with eigenvectors and
eigenvalues of polarization operator correspondingly. In the case of
strongly magnetized charge-symmetric plasma the eigenvalues of the
polarization operator can be obtained from the results of ~\cite{Shabad}:

\begin{eqnarray}
\P^{(1)}(q) &\simeq& - \frac{\alpha}{3 \pi}\,
q_{\mbox{\tiny $\bot$}}^2 -
q^2\, \Lambda(B), \label{P1}
\\[2mm]
{\cal P}^{(2)}(q) &\simeq& -\frac{2 eB \alpha}{\pi}\left[
H\left(\frac{q^2_{\mprl}}{4m^2} \right) + {\cal J}(\prl{q}) \right]
- q^2 \, \Lambda(B), \label{P2}
\\[2mm]
\P^{(3)}(q) &\simeq&  - \, q^2 \, \Lambda(B),
\label{P3}
\end{eqnarray}

where
$$
\Lambda(B) = \frac{\alpha}{3 \pi}\,\left[1.792 - \ln
(B/B_e)\right].
$$
\beq
\nonumber
{\cal J}(\prl{q}) = 4 \prl{q}^2 m^2 \int \frac{dp_z}{E} \,
\frac{f_E}
{(\prl{q}^2)^2 - 4 \prl{(pq)}^2}\,, \qquad E = \sqrt{p_z^2 + m^2},
\eeq
$f_E \, = \, [\exp{(E/T)} \, + \, 1]^{-1}$
is the electron distribution function,
\beq
\nonumber
H(z)=\frac{1}{\sqrt{z(1 - z)}} \arctan \sqrt{\frac{z}{1 - z}} - 1,
\ z \le 1.
\eeq

The four-vectors with indices $\bot$ and $\parallel$
belong to the Euclidean \{1, 2\}-subspace and the Minkowski
\{0, 3\}-subspace correspondingly in the frame were the magnetic
field is directed along third axis; $\varphi_{\alpha \beta} =  F_{\alpha
\beta} /B$ and
${\tilde \varphi}_{\alpha \beta} = \frac{1}{2} \varepsilon_{\alpha \beta
\mu \nu} \varphi_{\mu \nu}$ are the dimensionless field tensor and dual
field tensor correspondingly.
The tensors
$\Lambda_{\alpha \beta} = (\varphi \varphi)_{\alpha \beta}$,\,
$\widetilde \Lambda_{\alpha \beta} =
(\tilde \varphi \tilde \varphi)_{\alpha \beta}$, with equation
$\widetilde \Lambda_{\alpha \beta} - \Lambda_{\alpha \beta} =
g_{\alpha \beta} = diag(1, -1, -1, -1)$ are introduced.

The dispersion properties of the normal modes are defined from the
dispersion equations
\beq
q^2 - \P^{(\lm)}(q) = 0 \qquad (\lm = 1, 2, 3).
\label{disper}
\eeq
Their analysis shows that the modes
with $\lambda = 1, 2$ and with polarization vectors
\beq
\ee_\alpha^{(1)}(q) = \frac{(q \varphi)_\alpha}{\sqrt{q_{\mprp}^2}},
\qquad
\ee_\alpha^{(2)}(q) = \frac{(q \tilde \varphi)_\alpha}
{\sqrt{q_{\mprl}^2}}.
\label{epsilon}
\eeq
are only physical in the case under consideration, just as in the
pure magnetic field~\footnote{ Symbols 1 and 2 correspond to
the $\|$ and $\perp$ polarizations in pure magnetic
field~\cite{Adler:1971} and $E$- and $O$- modes in magnetized
plasma~\cite{Duncan:1995}.}. However, it should be emphasized that the
coincidence of the polarization vectors in the plasma is
approximate, to within $O(1/\beta)$.

\begin{figure*}
\centerline{\includegraphics{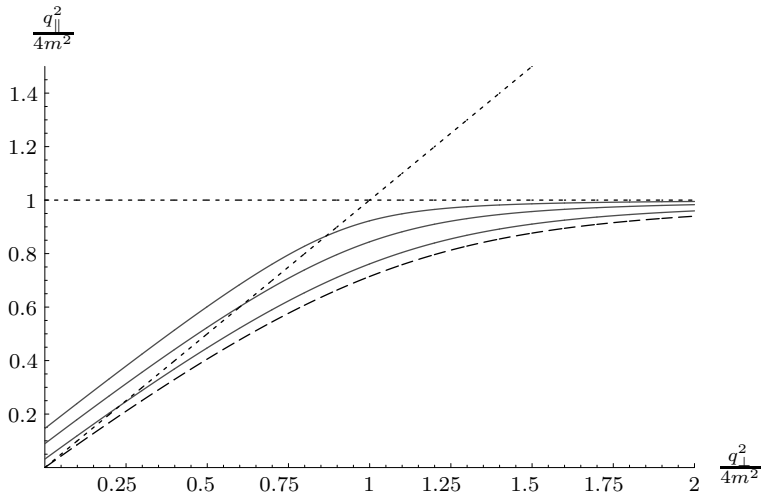}}
\vspace*{-2mm}
\caption{Photon dispersion laws in strong magnetic field $B/B_e = 200$
and neutral plasma vs. temperature:  $T = 1$ MeV
 (is upper curve), $T = 0.5$ MeV (is middle curve), $T = 0.25$ MeV
 (is lower curve).
Photon dispersion without plasma is depicted by dashed line.
Dotted line corresponds to the vacuum dispersion law, $q^2 = 0$. The angle 
between the photon momentum and the magnetic
field direction is $\pi/2$.
}
\label{fig:dis1}
\end{figure*}

\begin{figure*}
\centerline{\includegraphics{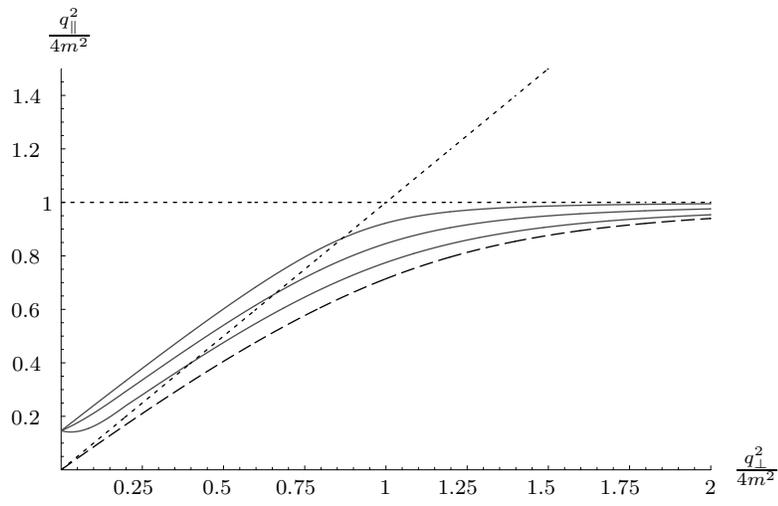}}
\vspace*{-2mm}
\caption{
Photon dispersion in a strong magnetic field ($B/B_e = 200$) and a charge-symmetric plasma
($T = 1$ MeV) at
various angles between the photon momentum and the magnetic
field direction: $\theta = \pi/2$ (upper solid curve), $\pi/6$ (middle
solid curve), and $\pi/12$ (lower solid curve). The dashed curve
represents the dispersion in the absence of a plasma.
}
\label{fig:dis2}
\end{figure*}

Notice, that in plasma only the eigenvalue $\P^{(2)}(q)$ modifies in
comparison with pure magnetic field case. It means that  the dispersion
law of the mode 1 is the same one as in the magnetized vacuum, where its
deviation from the vacuum law, $q^2 = 0$, is negligibly small.
From the other hand, the dispersion properties  of the mode 2 essentially differs from the
magnetized vacuum ones. In the Figure~\ref{fig:dis1} the photon dispersion in both
strong magnetic field and magnetized plasma at various temperatures are depicted.
One can see that in the presence of the magnetized plasma there exist the kinematical region, where $q^2 >0$ contrary 
to the case of  pure magnetic field. This fact could lead to the alteration of the different photon processes kinematic. 
For example, the photon splitting channel $\gamma _2 \to \gamma _1 \gamma_1$ forbidden in the magnetic field without plasma becomes allowed~\cite{RCh05}.
Another important distinction is an essentially
different dependence of the dispersion law in variables $q_{\mprl}^2$
and $q_{\mprp}^2$ on the angle between the photon momentum
and the magnetic field direction (see Fig.~\ref{fig:dis2}).

It follows from Eq.~(\ref{P2}) that the eigenvalue of the
polarization operator ${\cal P}^{(2)}$ becomes large near the electron-
positron pair production threshold, $q_{\mprl}^2=4 m^2$. This suggests that
the renormalization of the wave function for a photon
of this polarization should be taken into account:
\beq
\ee_{\alpha}^{(2)}(q) \to \ee_{\alpha}^{(2)}(q) \sqrt{Z_2}, \quad
Z^{-1}_2 = 1 - \frac{\partial {\cal P}^{(2)}(q)}{\partial \omega^2}.
\label{eps}
\eeq

To calculate the amplitude of process in strong magnetic field one
should use the Dirac equation solutions at the ground Landau level.
For the electron propagator it is relevant to use its asymptotic
form~\cite{KM_Book}. Then substituting the polarization
vectors~(\ref{epsilon}) one can obtain partial amplitudes for
different polarization configuration of the initial and final
photons. It is possible to present them in the covariant form:
\begin{eqnarray}
{\cal M}_{1 \to 1} = -\frac{8 \pi \alpha m}{eB}\,
\frac{(q \varphi q')(q \tilde \varphi q')}
{\sqrt{q^2_{\mprp} q'^2_{\mprp} (-Q^2_{\mprl})}},
\label{eq:M11}
\end{eqnarray}
\begin{eqnarray}
{\cal M}_{1 \to 2} = -\frac{8 \pi \alpha m}{eB}\,
\frac{(q \Lambda q')(q' \tilde \Lambda Q)}
{\sqrt{q^2_{\mprp} q'^2_{\mprl} (-Q^2_{\mprl})}},
\label{eq:M12}
\end{eqnarray}
\begin{eqnarray}
{\cal M}_{2 \to 1} = \frac{8 \pi \alpha m}{eB}\,
\frac{(q \Lambda q')(q \tilde \Lambda Q)}
{\sqrt{q^2_{\mprl} q'^2_{\mprp} (-Q^2_{\mprl})}},
\label{eq:M21}
\end{eqnarray}
\begin{eqnarray}
{\cal M}_{2 \to 2} = 16 i \pi \alpha m\, \frac{\sqrt{q^2_{\mprl}
q'^2_{\mprl}}\,\sqrt{(-Q^2_{\mprl})}\, \varkappa} {(q \tilde \Lambda
q')^2 - \varkappa^2 (q \tilde \varphi q')^2}, \label{eq:M22}
\end{eqnarray}

\noindent where $\varkappa = \sqrt{1 - 4m^2/Q^2_{\mprl}}$ and
$Q^2_{\mprl} = (q - q')^2_{\mprl} <0$. We can see from last equations that all 
amplitudes  except ${\cal M}_{2 \to 2}$ are suppressed by magnetic field
strength. Therefore one could expect that mode 2 has the largest scattering
absorption rate.

\section{Photon absorption rate and cross section in the strongly magnetized medium}
\label{Sec3}

\begin{figure*}
\centerline{\includegraphics{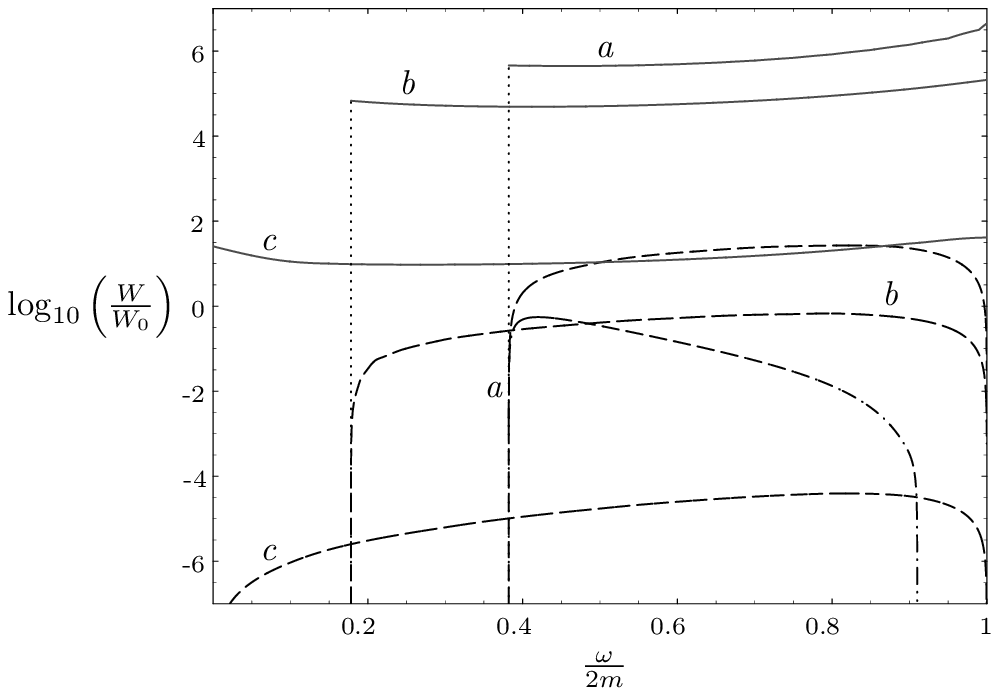}}
\vspace*{-2mm}
\caption{The dependence of the photon absorption rate  for  channels
$\gamma_2 e^{\pm} \to \gamma_2 e^{\pm}$ (solid line) and
$\gamma_2 e^{\pm} \to \gamma_1 e^{\pm}$ (dashed line) on energy of initial photon in 
strong magnetic field $B/B_e \, = \, 200$ at $T \,=\,1 MeV$ -- {\it a}, $T \,=\,250 keV$ -- {\it b} and
$T \,=\,50 keV$ -- {\it c}. The dotted line corresponds to the probability of photon splitting,
$\gamma_2 \to \gamma_1 \gamma_1$, at $T \, = \, 1 MeV$~\cite{RCh05}.
Here  $W_0 \, = \, (\alpha/\pi)^3\, m \simeq 3.25\cdot 10^2 cm^{-1}$.}
\label{fig:W22}
\end{figure*}

\begin{figure*}
\centerline{\includegraphics{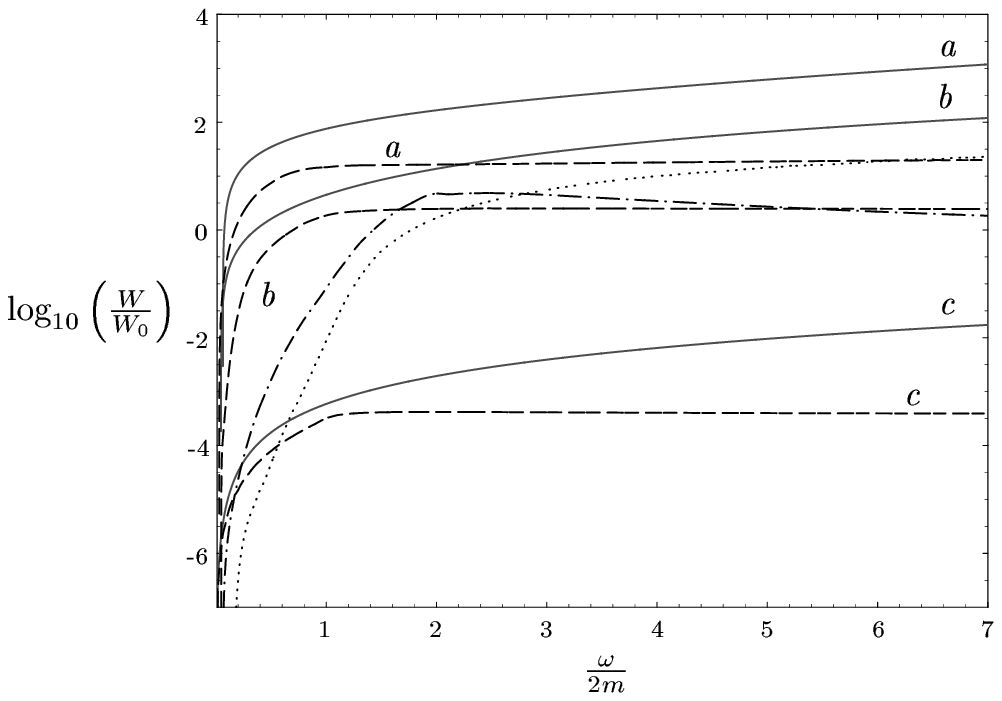}}
\vspace*{-2mm}
\caption{The dependence of the photon absorption rate  for  channels
$\gamma_1 e^{\pm} \to \gamma_1 e^{\pm}$ (solid line) and
$\gamma_1 e^{\pm} \to \gamma_2 e^{\pm}$ (dashed line) on energy of initial photon in 
strong magnetic field $B/B_e \, = \, 200$ at $T \,=\,1 MeV$ -- {\it a}, $T \,=\,250 keV$ -- {\it b} and
$T \,=\,50 keV$ -- {\it c}.
The dotted and chain lines correspond to the probability of photon splitting, 
$\gamma_1 \to \gamma_1 \gamma_2$ and $\gamma_1 \to \gamma_2 \gamma_2$, 
respectively, at $T \, = \, 50 keV $~\cite{RCh05}.
Here  $W_0 \, = \, (\alpha/\pi)^3\, m \simeq 3.25\cdot 10^2 cm^{-1}$.}
\label{fig:W11}
\end{figure*}

To analyse the efficiency of the process under consideration and to compare it 
with other competitive reactions we  calculate the photon absorption rate which
can be defined in the following way:
\begin{eqnarray}
&&W_{\lambda e^{\pm} \to \lambda^{'} e^{\pm}} = \frac{eB}{16 (2\pi)^4 \omega_{\lambda}}
\int \mid {\cal M_{\lambda \to \lambda^{'} }}\mid^2
Z_{\lambda}Z_{\lambda^{'}} \times
\label{eq:Wscatt}\\
&&\times f_{E}\, (1-f_{E'}) \, (1 + f_{\omega'})
\delta (\omega_{\lambda}({\bf k}) + E - \omega_{\lambda^{'}}({\bf k'}) - E')
\frac{dp_z\,d^3 k^{'}}{ E E' \omega_{\lambda^{'}}},
\nonumber
\end{eqnarray}
where $f_{\omega'} \, = \, [\exp{(\omega'/T)} \, - \, 1]^{-1}$ is the
photon distribution function. In the case of the rarefied plasma ($T \ll m$)
these absorption rates can be expressed in term of partial cross sections~\footnote{Hereafter the initial photon propagation across magnetic field is considered}
$W_{\lambda \to \lambda^{'}} = W_{\lambda e^{-} \to \lambda^{'} e^{-}} +
W_{\lambda e^{+} \to \lambda^{'} e^{+}} =
n_e \sigma_{\lambda \to \lambda^{'}}$:
\begin{eqnarray}
\sigma_{1 \to 1} = \frac{3}{8}\, \sigma_T  \left ( \frac{B_e}{B} \right )^2 \frac{\omega^2}{m^2} \,
\left [\frac{\omega + 2m}{2(\omega + m)} + \frac{m}{\omega}
\ln \left (1 + \frac{\omega}{m} \right)\right ],
\label{eq:sigma11}
\end{eqnarray}
\begin{eqnarray}
\sigma_{2 \to 1} = \frac{3}{8}\, \sigma_T \left (  \frac{B_e}{B}\right )^2 \frac{q^2_{\mprp}}{m^2} \, Z_2 \,
\left [\frac{\omega + 2m}{2(\omega + m)} - \frac{m}{\omega}
\ln \left (1 + \frac{\omega}{m} \right)\right ],
\label{eq:sigma21}
\end{eqnarray}
\begin{eqnarray}
\sigma_{1 \to 2} = \frac{3}{4}\, \sigma_T  \left (  \frac{B_e}{B} \right)^2
\frac{(\omega + 2m)^2}{\omega (\omega + m)} \!\!\!\! \int \limits_0^{\,\,\,\,\,\,\omega^2\!/4m^2} \!\!\!\!\!\!\! dz
\left(1 + \frac{3}{2} \xi \,\frac{H(z)}{z} \right)
\sqrt{\frac{\omega^2 - 4 m^2 z}{(\omega + 2m)^2 - 4 m^2 z}},
\label{eq:sigma12}
\end{eqnarray}
\begin{eqnarray}
\!\!\!\!\!\!\!\!\!\!\!\!\sigma_{2 \to 2}& = & 6\, \sigma_T \, \frac{m^4 (\omega + m)}
{\omega^3 (\omega + 2m)^2}\, Z_2 \,
\left [\frac{\omega(\omega + 2m)}{(\omega + m)(2m - \omega)} -
\ln \left (1 + \frac{\omega}{m} \right) + \right.
\nonumber
\\
&+&\left.\frac{2 \omega (\omega - m) (2m + \omega)}
{(\omega + m)(2m - \omega)\sqrt{4m^2 - \omega^2}}\,
\arctan{\frac{\omega}{\sqrt{4m^2 - \omega^2}}} \right ],
\label{eq:sigma22}
\end{eqnarray}
\noindent where $\sigma_T$ is the Thompson cross section, $\xi = \frac{\alpha}{3 \pi} \frac{B}{B_e}$ is the parameter characterizing magnetic field influence, $q^2_{\mprp}= \omega^2 - \P^{(2)}(\omega^2)$  and the number of  electron (positron) density in a
strongly magnetized, charge-symmetric rarefied plasma can be estimated as
\begin{eqnarray}
n_e \simeq eB \sqrt{\frac{mT}{2\pi^3}}\,e^{-m/T}.
\label{eq:ne}
\end{eqnarray}
In the low energy limit ($\omega \ll m$) the formulas (\ref{eq:sigma11}-\ref{eq:sigma22}) can be presented in the following form:
\begin{eqnarray}
\sigma_{1 \to 1} &=& \frac{3}{4}\, \sigma_T  \left ( \frac{B_e}{B} \right )^2 \frac{\omega^2}{m^2}, 
\quad
\sigma_{1 \to 2} = \frac{1}{4}\, \sigma_T  \left ( \frac{B_e}{B} \right )^2 \frac{\omega^2}{m^2}\, (1 + \xi),
\label{eq:sigma1sT}
\\
\sigma_{2 \to 1} &=& \frac{1}{16}\, \sigma_T  \left ( \frac{B_e}{B} \right )^2 \frac{\omega^2 (\omega^2-\omega_{pl}^2)}{m^4} \theta(\omega - \omega_{pl}) ,
\quad
\sigma_{2 \to 2} =  \frac{\sigma_T}{1+\xi},
\label{eq:sigma2sT}
\end{eqnarray}
where $\omega_{pl}$ is plasma frequency defined by equation $\omega_{pl}^2 - \P^{(2)}(\omega_{pl}, \vec k \to 0 )$ = 0.
One can see that the presence of magnetized plasma slightly influences on the process cross-sections in this limit. Moreover, the corrections connected with photon dispersion and wave function renormalization are significant only for  $\xi \sim 1$, i.e. when magnetic field is rather strong $B \sim 10^3 B_e$. In the case $\xi \ll 1$ which is relevant for the models of magnetar magnetosphere emission the formulas (\ref{eq:sigma1sT}, \ref{eq:sigma2sT}) coincide with the well-known result~\cite{Herold:1979}. It is interesting also to consider the case of  $\omega \sim 2 m$. In this region the magnetized vacuum influence is large and the cross-sections (\ref{eq:sigma11}-\ref{eq:sigma22}) have the form:
\begin{eqnarray}
\sigma_{1 \to 1} &=&  \sigma_T  \left ( \frac{B_e}{B} \right )^2 \, \Big [1 + \frac{3}{4} \ln 3 \Big ], 
\quad
\sigma_{1 \to 2} = 4\, \sigma_T  \left ( \frac{B_e}{B} \right )^2 \,\Big [1 - \frac{3}{4} \ln 3 \Big ]\, (1 + 2 \xi),
\label{eq:sigma12m}
\\
\sigma_{2 \to 1} &=& 2 \, \sigma_T  \left ( \frac{B_e}{B} \right )^2 \, \Big [1 - \frac{3}{4} \ln 3 \Big ]\, \left (1 - \frac{\omega^2}{4 m^2} \right ),
\quad
\sigma_{2 \to 2} =  \frac{\sigma_T}{2 \xi}.
\label{eq:sigma22m}
\end{eqnarray}
To investigate the Compton scattering under hot plasma conditions ($T \sim m$) we have made the numerical calculations of photon absorption  rates for the various channels. The results are represented in Figures \ref{fig:W22} and \ref{fig:W11}. 


\section{Discussion}
\label{Sec4}

In the models of soft gamma repeaters spectrum formation the dependence of photon absorption rates on energy and temperature plays an important role. It could influence on the shape of emergent spectrum and defines the temperature profile in the emission region during bursts in SGRs~\cite{Duncan:1995, Lyubarsky:2002}.  
In the both figures \ref{fig:W22} and \ref{fig:W11} one can see that photon absorption rates corresponding to Compton scattering are the fast increasing functions of temperature. At the same time, the channels with initial photon of mode 1 and mode 2 have different character of the absorption coefficient  energy dependence. As shown in the  Fig.\ref{fig:W22} the absorption rates for the reactions 
$\gamma_2 e^{\pm} \to \gamma_2 e^{\pm}$  and $\gamma_2 e^{\pm} \to \gamma_1 e^{\pm}$ have thresholds. They are  caused by mode 2 dispersion relation with plasma frequency and indicated the fact that the electromagnetic wave corresponding to mode 2 can't propagate with energy below $\omega_{pl}$. In the region above plasma frequency $W_{\gamma_2 e^{\pm} \to \gamma_2 e^{\pm}}$ is slightly depends on energy while $W_{\gamma_2 e^{\pm} \to \gamma_1 e^{\pm}}$ tends to zero in the vicinity of $\omega_{pl}$ and pair creation threshold, $\omega = 2m$. We would like to note that in the vicinity of pair creation threshold taking account of wave function renormalization and photon dispersion becomes very important and defines the processes rates dependence on energy, temperature and magnetic field. In the  region $\omega \ge 2 m$ the photon of mode 2 is unstable due to the  process $\gamma _2 \to e^+ e^-$ with rate much lager than Compton scattering.  The energy dependence of $W_{\gamma_1 e^{\pm} \to \gamma_1 e^{\pm}}$ and 
$W_{\gamma_1 e^{\pm} \to \gamma_2 e^{\pm}}$ is depicted in Fig.\ref{fig:W11}. On can see the fast increasing of absorption coefficients at low energies and rather slow dependence at $\omega \gtrsim 2 m$. 

The previous investigation of radiation transfer problem in strongly magnetized plasma have shown that along with Compton scattering process the photon splitting $\gamma \to \gamma \gamma$ could play a significant role as mechanism of photon production~\cite{Duncan:1995}. To compare these processes we depict the probabilities  of the photon splitting channels on the same plots (see dotted and chain lines in Fig.\ref{fig:W22} and Fig.\ref{fig:W11}). On can see that at temperature $T \sim m$  and in kinematical region $\omega \le 2 m $ the main photon splitting process is $\gamma_2 \to \gamma_1 \gamma_1$ forbidden in pure magnetic field~\cite{RCh05}. It is seen also that the rate of the process is much less than Compton scattering ones. Nevertheless,  it could be an effective photon production mechanism at temperatures under consideration. The probabilities of the channels $\gamma_1 \to \gamma_1 \gamma_2$ and $\gamma_1 \to \gamma_2 \gamma_2$ increase with temperature falling  and become comparable and even larger than Compton scattering rates $W_{\gamma_1 e^{\pm} \to \gamma_1 e^{\pm}}$ and  $W_{\gamma_1 e^{\pm} \to \gamma_2 e^{\pm}}$. As shown in Fig.~\ref{fig:W11} the process of photon splitting strongly dominates over Compton scattering at $T=50 keV$.
It was claimed previously that the effect of the cold strongly magnetized plasma on photon splitting is not pronounced and the vacuum approximation can be used in the most calculations~\cite{Bulik:1997,Elmfors:1998}. We can see now that in the presence of hot plasma the process of photon splitting could not be only intensive source of photon production but also an effective absorption mechanism. 

\begin{figure*}
\centerline{\includegraphics{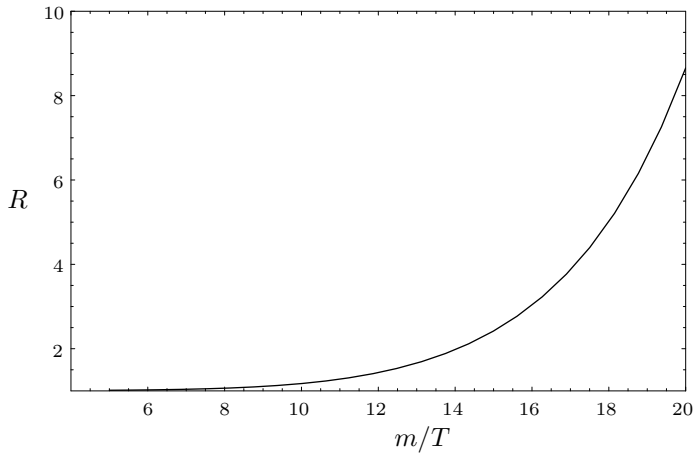}}
\vspace*{-2mm}
\caption{The ratio of the Rosseland mean free paths without and with 
taking into account of photon splitting process as a function of the 
inverse temperature at $B = 200 B_e$} 
\label{fig:RC_CSp}
\end{figure*}

Let us illustrate this  fact in the framework of the magnetar model of SGR burst. It is known that the radiation transfer in the magnetically trapped plasma may be described  as s diffusion of the 1-mode photons whereas 2-mode photons are locked~\cite{Duncan:1995,Lyubarsky:2002}. The last circumstance is connected especially with weak dependence of the 2-mode absorption coefficient on photon energy (see Fig.~\ref{fig:W22}). In the assumption that the properties of the radiation field do not change much over absorption length $W^{-1}$ the radiation transfer may be conveniently described in terms of the Rosseland mean free path of 1-mode photons~\cite{Duncan:1995,Lyubarsky:2002}
\begin{equation}
 \lambda_R = \frac{\int W^{-1} \omega^4 \exp(\omega/T) [\exp(\omega/T)-1]^{-2} d \omega}{\int \omega^4 \exp(\omega/T) [\exp(\omega/T)-1]^{-2} d \omega},
\label{eq:lR}
\end{equation}
where $W$ is the total absorption coefficient. Then the temperature distribution in the magnetically trapped plasma region could be defined from the diffusion equation:
\begin{equation}
 F(z)= \frac{16 \lambda_R \, \sigma}{3}\,T^3 \, \frac{\partial T}{\partial z},
\label{eq:Difuss} 
\end{equation}
where $\sigma$ is the Stefan-Boltzmann constant, $F(z)$ is the radiation flux. 
If only Compton scattering process is considered then
\begin{equation}
 W = W_{\gamma_1 e^{\pm} \to \gamma_1 e^{\pm}} + W_{\gamma_1 e^{\pm} \to \gamma_2 e^{\pm}}.
\label{eq:WCompt}
\end{equation}
To take into account the effect of photon splitting one should add  term 
$W_{\gamma_1 \to \gamma_1 \gamma_2} + W_{\gamma_1 \to \gamma_2 \gamma_2}$ to the equation (\ref{eq:WCompt}). In the figure~\ref{fig:RC_CSp} the ratio of the Rosseland mean free path for only Compton scattering to  Rosseland mean free path  with photon splitting is depicted. One can see that at low temperatures the additional absorption process of photon splitting leads to the significant decreasing of the Rosseland mean free path. The solution of the diffusion equation (\ref{eq:Difuss}) is out of the scope of this article. However, we would like to note that the more detail analysis of the radiation transfer needs the consistent solution of Boltzmann equation for the photon occupation number and radiative transfer equation in the wide range of temperatures 
($ 10 keV \lesssim T \lesssim 1 MeV$). 

In conclusion, we have investigated the influence of the strongly 
magnetized hot plasma on the Compton scattering process with taking into 
account of photon dispersion and large radiative corrections. It was found 
that absorption rate of the dominant scattering channel of the 2-mode 
photon weakly depends on energy in  wide temperature range whereas the 
1-mode absorption coefficient is fast increasing function of photon 
energy. The comparison of the photon splitting effect and Compton 
scattering shows that the influences of these reaction on the 1-mode 
radiation transfer are competitive in rarefied plasma $(T \ll m)$.

\medskip

\noindent{\bf Acknowledgements}  

We express our deep gratitude to the organizers of the 
Seminar ``Quarks-2006'' for warm hospitality.
This work supported in part by the Council on Grants by the President 
of Russian Federation for the Support of Young Russian Scientists and 
Leading Scientific Schools of Russian Federation under the Grant 
No. NSh-6376.2006.2 and No. MK-1097.2005.2 (MC), and by the Russian Foundation for Basic Research 
under the Grant No. 04-02-16253. Work of MC is supported
by the DAAD (Deutscher Akademischer Austauchdienst).


%
\end{document}